\documentclass[showpacs,pre,amsmath,twocolumn]{revtex4}
\usepackage{graphicx}
\usepackage{amsmath,amsfonts,amssymb}
\begin{document}
\title{Langevin dynamics of the Coulomb frustrated ferromagnet:
 a mode-coupling  analysis.}

\author{M. Grousson$^{1,2}$, V. Krakoviack$^{1,3}$, G. Tarjus$^1$ and
P. Viot$^1$}
\affiliation{$^1$
Laboratoire de Physique  Th{\'e}orique des Liquides\\ Universit{\'e} Pierre et
Marie Curie 4, place Jussieu, 75252 Paris Cedex 05, France}
\affiliation{$^2$
Department of Chemistry and Biochemestry, \\
University of California, 
Los Angeles, CA, 90095, USA 
}

\affiliation{$^3$ CUC\textsuperscript{3},
Department of Chemistry, \\ 
Lensfield Road, 
Cambridge CB2 1EW, 
United Kingdom 
}

\begin{abstract}
We study the Langevin dynamics  of the soft-spin, continuum version of
the Coulomb  frustrated   Ising ferromagnet.  By  using  the dynamical
mode-coupling approximation, supplemented by reasonable approximations
for describing the equilibrium  static  correlation function, and  the
somewhat improved  dynamical self-consistent screening  approximation,
we find that the  system displays  a  transition from an ergodic  to a
non-ergodic behavior.  This transition is similar  to that obtained in
the idealized mode-coupling theory of glassforming  liquids and in the
mean-field   generalized spin glasses   with one-step replica symmetry
breaking. The   significance   of this   result  and  the  relation to  the
appearance of a complex free-energy landscape are also discussed.
\end{abstract}
\maketitle               
\section{Introduction}\label{sec:introduction}
Models with  a competition between  a short-range ordering interaction
and a long-range frustrating interaction have been recently introduced
to  explain the slowing down  of relaxation in supercooled liquids and
the resulting  glass transition\cite{KKZNT95}.   The  underlying
picture is as follows: any given  liquid possesses a locally preferred
structure   which is different   than  that of  the actual crystalline
phase, and   this  local arrangement of  the  molecules  in the liquid
cannot propagate at long distances to tile the whole space and form an
``ideal  crystal''  because of  ubiquitous frustration.   It  has been
argued\cite{KKZNT95} that, for   weak enough  frustration,  this
phenomenon can be described via effective  interactions acting on very
different length scales: a short-range term describing the tendency to
extend the locally preferred  structure and a long-range, Coulomb-like
term  describing the  frustration-induced free-energy  cost associated
with  this spatial extension.    Such Coulomb frustrated systems  have
been shown,  both by scaling arguments\cite{KKZNT95,VTK00}
and by Monte Carlo simulation\cite{GrTV01}, to display the
generic features   observed  in  fragile glassforming   liquids,  most
notably the super-Arrhenius temperature  dependence of  the relaxation
time   and the  two-step,  non-exponential  decay  of the  correlation
function.

These Coulomb frustrated models have also been used in quite different
contexts to  describe the formation of  modulated spatial  patterns on
mesoscopic length scales, such as lamellar and cubic phases in diblock
copolymer        melts\cite{OK86,L80,FH87},        microemulsions       in
water-oil-surfactant mixtures\cite{S83,WCS92}, or  stripe phases
in high  temperature superconductors\cite{EK93}.   In all these
cases, slow relaxation  is usually observed,  and it has been recently
argued\cite{ScWo00} that high-temperature superconductors could indeed
form  a ``stripe glass'' in  which glassiness is self generated, i.e.,
does not  result from the presence of  quenched disorder.  This latter
result has been obtained through  an investigation of the properties of
the free-energy landscape of the  Coulomb frustrated $\phi^4$ scalar field
theory: by using a thermodynamic approach combining the replica method
proposed for  the  study  of structural  glasses\cite{M95,MP99} and  a
particular approximation, the self-consistent screening  approximation
(SCSA)\cite{AJBa74}, for calculating  the pair correlation  functions,
Schmalian and  Wolynes\cite{ScWo00} have  derived that the free-energy
landscape of  the  Coulomb frustrated model  becomes non-trivial below
some temperature $T_A$ at   which  an exponentially large number   of
metastable  states  appears;  the  associated  configurational entropy
decreases with further  decrease of the temperature  and vanishes at a
lower  temperature   $T_K$\cite{ScWo00,SW201,endnote1}. 

Motivated by these results  giving  evidence for fragile  glassforming
behavior in Coulomb frustrated   models, we have studied the  Langevin
dynamics of the  Coulomb frustrated $\phi^4$  scalar  field theory within
the  mode-coupling    and    related   approximations.   Mode-coupling
approaches  have   been      widely used   to   study      glassforming
liquids\cite{gotze91,Cummins99},  and        the  dynamical
ergodicity-breaking singularity predicted   to  occur in  the   weakly
supercooled liquid regions, albeit    ``avoided'' in real   systems, is
taken   by  many as   a   canonical feature  of  fragile  glassforming
systems.  It is  therefore  tempting  to investigate whether   Coulomb
frustrated models  also display this feature. 

The paper is  organized as follows.  We  first  present the model  and
summarize the equilibrium phase   behavior and the results  previously
obtained by Schmalian and Wolynes.    We also introduce the   Langevin
equation describing  the relaxational  dynamics   of the  system.   In
section III,  we  derive   the  evolution equations followed   by  the
equilibrium time-dependent  correlation function  obtained within  two
resummation schemes   of perturbative expansions:   the  mode-coupling
approximation  and the dynamical  SCSA.  Section IV  is devoted to the
search for an ergodicity-breaking   transition.  We find that  such  a
phenomenon is indeed observed  with the two approximations considered.
We also show that the dynamical singularity predicted by the dynamical
SCSA  coincides with   the temperature  $T_A$  at  which the   replica
analysis of Refs\cite{ScWo00,SW201} predicts the occurrence of an
exponentially large number  of  metastable states.   In section  V, we
present the full numerical  solution  of the mode-coupling  equations,
thereby obtaining  the time-evolution of  the equilibrium  correlation
function; this latter  is  similar to that  obtained  in the idealized
mode-coupling  theory  of  supercooled  liquids\cite{gotze91}  and  in
mean-field generalized spin-glass models\cite{KT87,BCKM96,CHS93}. In
the last  two section, we address  the question  of sensitivity of the
results to the level  and the details of  the approximation scheme and
we give some concluding remarks.

\section{Model}\label{sec:model}
We  consider  the  field-theoretical  version  of the  $3$-dimensional
Coulomb frustrated Ising ferromagnet defined by the  Hamiltonian
\begin{widetext}
\begin{align}\label{eq:1}
H[\phi]=&\frac{1}{2}\int
d^3{\bf{x}}\left\{(\nabla\phi({\bf{x}}))^2+r_0\phi^2({\bf{x}})+\frac{u}{2}\phi^4({\bf{x}})\right\}
+\frac{Q}{8\pi}\int                                                   d^3{\bf{x}}\int
d^3{\bf{x}}'\frac{\phi({\bf{x}})\phi({\bf{x'}})}{|{\bf{x}}-{\bf{x'}}|}\\\label{eq:2}
=&\frac{V}{2}\int                                         \frac{d^3{\bf
k}}{(2\pi)^3}\left(r_0+k^2+\frac{Q}{k^2}\right)\phi_{-{\bf   k}} \phi_{{\bf   k}}      +\frac{uV}{4}\int\frac{d^3{\bf
k_1}}{(2\pi)^3}\int\frac{d^3{\bf            k_2}}{(2\pi)^3}\int\frac{d^3{\bf
k_3}}{(2\pi)^3} \phi_{{\bf   k_1}}\phi_{{\bf     k_2}}\phi_{{\bf    k_3}}\phi_{-{\bf
k_1}-{\bf k_2}-{\bf k_3}},
\end{align}
\end{widetext}
where   $\phi({\bf{x}})$  is a   real  scalar   field ($\phi_{\bf k}$,   the
associated ${\bf k}-$Fourier component),  $V$ is the  volume, $u$ is a
strictly positive coupling constant, $Q$ is the frustration parameter,
and all momentum integrations are performed up to a cut-off $\Lambda$, i.e.,
$|{\bf k}|\leq \Lambda$;   $r_0$   is a  temperature-dependent  mass   which is
proportional to the   deviation, $T-T^0_{c,MF}$,  from  the mean-field
transition temperature of the unfrustrated $(Q=0)$ model.

\begin{figure}
\centering

\resizebox{8cm}{!}{\includegraphics{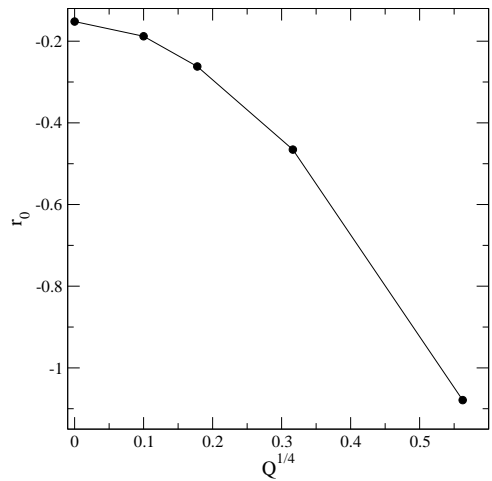}}
\caption{Temperature ($r_0$)-frustration ($Q^{1/4}$) equilibrium phase
diagram in the  self-consistent  Hartree approximation. The full  line
denotes the  fluctuation-induced first-order  transition  to modulated
phases; below this line, the  paramagnetic phase exists in a metastable
state. The coupling constant  $u$ is set  equal to  $1$.}\label{fig:1}

\end{figure}

The equilibrium partition function is 
\begin{equation}\label{eq:3}
Z=\int{\cal D}\phi  e^{-H(\phi)/T}.
\end{equation}
In what follows, we take $\Lambda=1$, and $T$ is set equal to $1$ in
Eq.~(\ref{eq:3}) so that the  whole temperature dependence is
contained in $r_0$. We are interested in the weak-frustration region for
which $Q<<1$.

In the  absence  of    frustration  $(Q=0)$,  the  model   defined  by
Eqs.~(\ref{eq:1}-\ref{eq:3})  reduces   to  the  usual $\phi^4$
theory. It undergoes a second-order transition at a finite temperature
$T^0_c$ to a broken-symmetry phase  characterized by a non-zero  value
of $<\phi_{{\bf k=0}}>$, where $<>$  denotes an equilibrium average.  For
$Q>0$, an  ordered phase with $<\phi_{{\bf k=0}}>\neq  0$ is  forbidden, but
the  system can  still undergo  a  phase  transition at  a temperature
$T_{DO}(Q)$ to  a   phase with  long-range   modulated  order.    This
transition has been studied by Monte Carlo simulation  for the case of
the Coulomb      frustrated  Ising    ferromagnet   on     a     cubic
lattice\cite{GTV01b} and via the self-consistent Hartree approximation
for a Hamiltonian similar   to Eq.~(\ref{eq:1}) describing  microphase
separation in diblock  copolymer  melts\cite{FH87}: it has then   been
shown that, whereas  the   mean-field theory predicts a   second-order
transition, the fluctuations change the   order of the transition  and
induce  a     first-order transition.    Such   a  fluctuation-induced
first-order transition was first discussed by Brazovskii for a related
model\cite{B75}. Within the self-consistent Hartree approximation, the
equilibrium (connected) correlation function
\begin{equation}\label{eq:4}
VC({\bf k})=<\phi_{-{\bf k}}\phi_{{\bf k}}>-<\phi_{-{\bf k}}><\phi_{{\bf k}}>
\end{equation}
is obtained via a self-consistent equation; for instance,  in the paramagnetic
phase where $<\phi_{{\bf k}}>=0$, this equation is
\begin{align}\label{eq:5}
C^{-1}({\bf{k}})=r_0+k^2+\frac{{Q}}{k^2}
+3u\int  \frac{d^3{\bf
q}}{(2\pi)^3}C({\bf{q}}).
\end{align}
The         renormalized   mass,     $r=r_0+3u\int         \frac{d^3{\bf
q}}{(2\pi)^3}C({\bf{q}})$, is then given by
\begin{equation}\label{eq:6}
r=r_0+3u\int  \frac{d^3{\bf
q}}{(2\pi)^3}\frac{1}{r+q^2+\frac{Q}{q^2}}.
\end{equation}
Since $k^2+\frac{Q}{k^2}$ is
minimum for non-zero wave  vectors with modulus $k_m=Q^{1/4}$, a value
characterizing the incipient  modulated order, one easily  checks that
$r$ only goes to zero when $r_0\to-\infty $. This means that the paramagnetic
phase is (meta)stable  at  all finite  ``temperatures'', its  spinodal being
depressed  to  $r_0\to-\infty $.   The Hartree   approximation  allows one to
calculate the free energy  of the paramagnetic phase  and that of  the
phase  with  modulated  order.   One then   obtains  the  temperature
$r_0^{DO}(Q)$ of the first-order transition at the  point at which the
two free energies are equal.  The details are given  in Appendix A and
the resulting phase diagram is shown in Fig.~\ref{fig:1}.

In their recent  work, Schmalian and Wolynes\cite{ScWo00} have applied
the   thermodynamic  approach   of  non-random   glassforming  systems
developed by M{\'e}zard   and Parisi\cite{MP99} to the Coulomb  frustrated
model. The basic idea, originally motivated by the behavior of a class
of  mean-field  generalized spin glasses  such  as  $p-$spin and Potts
glasses\cite{KT87,BCKM96},  is that glassiness  arises  because of the
occurrence of an exponentially large number of metastable states. This
occurrence, and  the associated emergence  of a non-zero complexity or
configurational  entropy, can be  more conveniently studied within the
replica  formalism\cite{M95}.  Approximations are  of course necessary
to  solve the corresponding many-body problem  and obtain all relevant
correlation  functions. Schmalian and   Wolynes have shown that within
the  SCSA\cite{AJBa74}, an approximation that  goes beyond the Hartree
result in  that   it   explicitly  includes  more   diagrams of    the
perturbative expansion\cite{endnote2}, there is a
temperature $T_A$ at which an exponentially large number of metastable
states emerges, as   signaled by a  non-zero configurational  entropy.
The  configurational  entropy  decreases  with  further decay  of  the
temperature and it vanishes at a temperature $T_K$ at which the system
undergoes   a    random   first-order   transition   to    an  ``ideal
glass''\cite{ScWo00}.

In the present work, we focus on the dynamics of the Coulomb
frustrated model defined by Eq.~(\ref{eq:1}). The starting point is
the Langevin equation,
\begin{equation}\label{eq:7}
\frac{\partial\phi_{{\bf{k}}}(t)}{\partial t}=-\frac{{\delta}(H[\phi_{{\bf{k}}}(t)]/V)}{\delta\phi_{-{\bf{k}}}(t)}+\eta_{{\bf{k}}}(t),
\end{equation}
that describes    the  purely relaxational dynamics   of   the system;
$\eta_{{\bf{k}}}(t)$       is   a      gaussian thermal      noise   with
$<\eta_{{\bf{k}}}(t)>=0$                                              and
$<\eta_{{\bf{k}}}(t)\eta_{{\bf{k}}'}(t')>=2T\delta({\bf{k}}+{\bf{k}}')\delta(t-t')$.
Eq.~(\ref{eq:7}) can be explicitly written as

\begin{align}\label{eq:8}
\frac{\partial\phi_{{\bf{k}}}(t)}{\partial t}&=-(r_0+k^2+\frac{{Q}}{k^2})\phi_{{\bf{k}}}\nonumber\\
-&u\int\frac{d^3{\bf{k}}_1}{(2\pi)^3}\int\frac{d^3{\bf{k}}_2}{(2\pi)^3}\phi_{{\bf{k}}_1}\phi_{{\bf{k}}_2}\phi_{{\bf{k}}
-{{\bf{k}}_1}-{{\bf{k}}_2}}+\eta_{{\bf{k}}}(t).
\end{align}

Solving this set   of  coupled non-linear   dynamical  equations is  a
daunting task,  and virtually all   available approximations amount to
performing    some   self-consistent   resummation    of  perturbative
expansions, e.g,. expansions in powers of  the coupling constant $u$ or
of the inverse of the number of components of the field, $1/n$, for an
$O(n)$  model.    In what   follows,  we    shall consider  two   such
self-consistent  resummation schemes, the mode-coupling  approximation
and the dynamical  SCSA\cite{endnote3}.

\section{Dynamical self-consistent  approximations}
To introduce the mode-coupling approximation, we first define the
time-dependent correlation function $C({\bf k},t,t')$ and the
associated response function $G({\bf k},t,t')$:

\begin{eqnarray}\label{eq:9}
\delta({\bf{k}}+{\bf{k}}')C({\bf{k}},t,t')&=&<\phi_{{\bf{k}}}(t)\phi_{{\bf{k}}'}(t')>,\\\label{eq:10}
\delta({\bf{k}}+{\bf{k}}')G({\bf{k}},t,t')&=&<\frac{\partial\phi_{{\bf{k}}}(t)}{\partial\eta_{{\bf{k}}'}(t')}>\nonumber\\
&=&\frac{1}{2T}<\phi_{{\bf{k}}}(t)\eta_{{\bf{k}}'}(t')>.
\end{eqnarray}
As in the preceding section, we  set $T=1$ in the following.

The perturbative expansion of $C({\bf k},t,t')$ and $G({\bf k},t,t')$
in powers of $u$ is more conveniently expressed by introducing the
zeroth-order correlation and response functions,
\begin{align}\label{eq:11}
C_0({\bf{k}},t,t')=&\frac{1}{\mu({\bf{k}})}\exp\left[\mu({\bf{k}})(t-t')\right],\\\label{eq:12}
G_0({\bf{k}},t,t')=&\exp\left[\mu({\bf{k}})(t-t')\right],
\end{align}
where
\begin{equation}\label{eq:13}
\mu({\bf{k}})=r_0+k^2+\frac{{{Q}}}{k^2},
\end{equation}
 and the two kernels
$\Sigma({\bf{k}},t,t')$ and $D({\bf{k}},t,t')$ defined though the standard
Dyson equations,
\begin{widetext}
\begin{eqnarray}\label{eq:14}
C({\bf{k}},t,t')&=&\int_{0}^{t}dt_1\int_{0}^{t'}dt_2G({\bf{k}},t,t_1)\left[2\delta(t_1-t_2)+D({\bf{k}},t_1,t_2)\right]G({\bf{k}},t',t_2),\\\label{eq:15}
G({\bf{k}},t,t')&=&G_0({\bf{k}},t,t')+\int_{t'}^{t}dt_1\int_{t'}^{t_1}dt_2G_0({\bf{k}},t,t_1)\Sigma({\bf{k}},t_1,t_2)G({\bf{k}},t_2,t').
\end{eqnarray} 
\end{widetext}

The diagrammatic representation of the perturbative expansion and a
detailed derivation of the mode-coupling approximation can be found in
Ref.\cite{BCKM96}; the only difference with the cases considered in
\cite{BCKM96} is the presence of the frustration term $Q/k^2$ in the
expression of $\mu(k)$, and we merely sketch here the  main steps of the
derivation.

The mode-coupling   approximation  amounts to expanding   the  kernels
$D({\bf{k}},t,t')$ and $\Sigma({\bf{k}},t,t')$ to   second order in  $u$
and replacing the  bare (zeroth-order)  functions $C_0({\bf{k}},t,t')$
and $G_0({\bf{k}},t,t')$ that  appear in the resulting  expressions by
their    renormalized       counterparts    $C({\bf{k}},t,t')$     and
$G({\bf{k}},t,t')$. This leads to\cite{BCKM96}:
\begin{widetext}
\begin{eqnarray}\label{eq:16}
D({\bf{k}},t,t')\simeq6u^2\int \frac{d^3{\bf k}_1}{(2\pi)^3}\int \frac{d^3{\bf k}_2}{(2\pi)^3}
C({\bf{k}}_1,t,t')
C({\bf{k}}_2,t,t')C({\bf k}-{\bf k }_1-{\bf k }_2,t,t'),\\ \label{eq:17}
\Sigma({\bf{k}},t,t')\simeq18u^2\int \frac{d^3{\bf k}_1}{(2\pi)^3}\int \frac{d^3{\bf
k}_2}{(2\pi)^3}
C({\bf{k}}_1,t,t')C({\bf{k}}_2,t,t')G({\bf k}-{\bf k }_1-{\bf k }_2,t,t').
\end{eqnarray} 
\end{widetext}

At  the    same time  $\mu({\bf{k}})$  is  renormalized  to  include the
so-called tadpole  diagrams,  which  replaces Eq.~(\ref{eq:13}) by  an
expression  similar to   that   obtained  within the  static   Hartree
approximation.

In this work, we are interested by the dynamical properties of the
system {\it at equilibrium}: therefore, the fluctuation-dissipation
theorem and the time-translation invariance apply, which reduces the
dependence upon the two times $t$ and $t'$ to the mere dependence on
the difference $t-t'$ and gives
\begin{equation}\label{eq:18}
G({\bf{k}},t)=-\Theta(t)\frac{\partial C({\bf{k}},t)}{\partial t}
\end{equation} 
and 
\begin{equation}\label{eq:19}
\Sigma({\bf{k}},t)=-\Theta(t)\frac{\partial D({\bf{k}},t)}{\partial t},
\end{equation}    
where  $\Theta(t)$ is the Heaviside step  function ($\Theta(t)$  is equal to $0$
for  $t<0$  and  to  $1$     pour $t>0$).   Applying   the    operator
$G_0^{-1}({\bf k})=\left[\mu({\bf{k}})+\frac{\partial}{\partial t}   \right]$ to both  sides of
Eq.~(\ref{eq:15}) yields
\begin{equation}\label{eq:20}
\frac{\partial G({\bf{k}},t)}{\partial t}=\delta(t)-\mu({\bf{k}}) G({\bf{k}},t)+\int_{0}^tdt'\Sigma({\bf{k}},t-t')G({\bf{k}},t'),
\end{equation} 
which, when combined with the time derivative of Eq.~(\ref{eq:18}), 
\begin{equation}\label{eq:21}
 \frac{\partial G({\bf{k}},t)}{\partial t}=-\delta(t)\left(\frac{\partial C({\bf{k}},t)}{\partial t}\right)_{t=0}-\Theta(t)\frac{\partial^2 C({\bf{k}},t)}{\partial t^2}, 
\end{equation}  
 gives
\begin{equation}\label{eq:22}
\left(\frac{\partial C({\bf{k}},t)}{\partial t}\right)_{t=0}=-1.
\end{equation}  
For $t>0$, the equation for the response function thus reads
\begin{equation}\label{eq:23}
\frac{\partial G({\bf{k}},t)}{\partial t}=-\mu({\bf{k}}) G({\bf{k}},t)+\int_{0}^tdt'\Sigma({\bf{k}},t-t')G({\bf{k}},t')
\end{equation} 
with  the    initial   condition  $G({\bf{k}},t=0^+)=1$.    By Laplace
transforming Eqs.~(\ref{eq:18}),   (\ref{eq:19}) and (\ref{eq:23}) and
using  the      initial   condition  $\left(\frac{\partial  C({\bf{k}},t)}{\partial
t}\right)_{t=0}=-G({\bf{k}},t=0^+)=-1$, one finally obtains
\begin{align}\label{eq:24}
1=&\left(z{\hat{C}}({\bf{k}},z)+C({\bf{k}},t=0)\right)\times \nonumber\\\times& \left(\mu({\bf{k}})-iz-z{\hat{D}}({\bf{k}},z)-D({\bf{k}},t=0) \right),
\end{align}
where $\hat{C}({\bf{k}},z)=i\int_{0}^{\infty}dte^{izt}C({\bf{k}},t)$, and  a
similar expression holds for ${\hat{D}}({\bf{k}},z)$. Going back to
the time dependence leads to 
\begin{align}\label{eq:25}
\frac{\partial C({\bf{k}},t)}{\partial t}=&- \left( \mu({\bf{k}})-D({\bf{k}},t=0)
\right)C({\bf{k}},t)\nonumber\\
-&\int_0^tdt'D({\bf{k}},t-t')\frac{\partial C({\bf{k}},t')}{\partial t'}
\end{align}
with the initial condition $C({\bf{k}},t=0)= \left(
\mu({\bf{k}})-D({\bf{k}},t=0)\right)^{-1}$ that follows from
Eqs.~(\ref{eq:22}) and (\ref{eq:25}).

The  mode-coupling approximation  finally   results in   the following
self-consistent equation for  the  time-dependent correlation function
at equilibrium:
\begin{align}\label{eq:26}
\frac{\partial C({\bf{k}},t)}{\partial t}=&-
C({\bf{k}},t=0)^{-1}C({\bf{k}},t)\nonumber\\
-&\int_0^tdt'D({\bf{k}},t-t')\frac{\partial C({\bf{k}},t')}{\partial t'}
\end{align}

with
\begin{align}\label{eq:27}
D({\bf{k}},t)=&6u^2\int \frac{d^3{\bf k}_1}{(2\pi)^3}\int \frac{d^3{\bf
k}_2}{(2\pi)^3}C({\bf{k}}_1,t)C({\bf{k}}_2,t)\times\nonumber
\\\times &C({\bf k}-{\bf k }_1-{\bf k }_2,t).
\end{align}

Except  for the absence  of inertial term, $\frac{\partial^2 C({\bf{k}},t)}{\partial
t^2}$, in   the  purely  relaxational  dynamics   associated with  the
Langevin   equation  and the  cubic  dependence  of  the memory kernel
$D({\bf{k}},t)$ on the correlation function $C({\bf{k}},t)$, the above
equations are similar to the  mode-coupling equations used to describe
the        time-dependent     density  fluctuations    in  supercooled
liquids\cite{gotze91}; they are also  analogous  to those derived  for
the mean-field spin glass with $4$-spin interactions\cite{BCKM96,CHS93}.

The necessary input for solving the self-consistent Eqs.~(\ref{eq:26})
and  (\ref{eq:27})   is  the   knowledge  of  the   equilibrium static
correlation function  $C({\bf{k}},t=0)$. Treating the statics  and the
dynamics of  the system on an  equal footing, as  for instance done in
the  above  derivation,    leads  to   considering   a   mode-coupling
approximation for  the static  correlation function, $C({\bf{k}},t=0)=
\left( \mu({\bf{k}})-D({\bf{k}},t=0)\right)^{-1}$. However,   we   shall
rather  introduce  more  flexibility in  the   mode-coupling scheme (a
flexibility that   goes   with   the   many  ways  to  implement   the
self-consistency at the second order of the perturbative expansion) by
allowing $C({\bf{k}},t=0)$ to be computed with several approximations,
such  as the Hartree approximation   and the SCSA,  that  are a priori
better  behaved than the  mode-coupling approximation  as  far  as the
static properties are concerned.

A  somewhat refined  resummation  scheme is provided  by the dynamical
SCSA.  (As mentioned in \cite{endnote2}, it  consists in using an
$n$-component vector  field  $\phi$, resumming self-consistently  all the
diagrams of  order $1/n$ in the  large $n$ expansion, and, eventually,
for the problem considered here,  setting $n$ equal  to $1$.)  Details
on the  derivation of the  approximate equation for the time-dependent
correlation  function     $C({\bf{k}},t)$      can    be   found    in
Refs.\cite{BCKM96,MCJPB97}.  A  convenient    way to  proceed   is  to
introduce a  complex auxiliary  field  $\sigma({\bf{x}})$ such that the  partition
function, Eq.~(\ref{eq:3}), can be rewritten $Z=\int \int{\cal D}\phi {\cal D}\sigma
e^{-H[\phi,\sigma]}$ (here $T=1$) with
\begin{align} H[\phi  ,\sigma]=&\frac{1}{2}\int
d^3{\bf{x}}\left\{(\nabla\phi({\bf{x}}))^2+r_0\phi^2({\bf{x}})-\sigma^2({\bf{x}})\right.\nonumber\\
+&\left.\sqrt{2u}\sigma({\bf{x}})\phi^2({\bf{x}})\right\}\nonumber\\   +&\frac{Q}{8\pi}\int
d^3{\bf{x}}\int d^3{\bf{x}}'\frac{\phi({\bf{x}})\phi({\bf{x'}})}{|{\bf{x}}-{\bf{x'}}|}.
\end{align}
The      field       $\sigma({\bf{x}})$          is         such       that
$<\sigma({\bf{x}})>=\sqrt{u/2}<\phi^2({\bf{x}})>$ and    its    connected pair
correlation                                                   function
$C_\sigma({\bf{x}},{\bf{x}}')=<\sigma({\bf{x}})\sigma({\bf{x}}')>_c$ is equal to
\begin{align}
C_\sigma({\bf{x}},{\bf{x}}')=&-\delta({\bf{x}}-{\bf{x}}')\nonumber\\
+&u/2(<\phi^2({\bf{x}})\phi^2({\bf{x}}')>-<\phi^2({\bf{x}})><\phi^2({\bf{x}}')>).
\end{align}
One can then apply to the dynamics of the coupled fields $\sigma({\bf{x}})$
and  $\phi({\bf{x}})$   a treatment similar      to that sketched  above.
Defining the equilibrium time-dependent correlation function $C_\sigma({\bf
k},t)$                                                             via
$\delta({\bf{k}}+{\bf{k}}')C_\sigma({\bf{k}},t)= <\sigma_{{\bf{k}}}(t)\sigma_{{\bf{k}}'}(0)>_c$,
the associated  kernel $D_\sigma( {\bf{k}},t)$  obtained through  the Dyson
equation  (see  above), and    similar   functions for the    response
properties,  one  can  perform a  mode-coupling  approximation  to the
coupled   dynamical equations   for    the  fields  $\sigma({\bf{x}})$  and
$\phi({\bf{x}})$. This leads to Eq.~(\ref{eq:26}) with $D({\bf{k}},t)$ now
given by\cite{MCJPB97}
\begin{equation}\label{eq:28}
D({\bf{k}},t)=2u\int \frac{d^3{\bf p}}{(2\pi)^3}C_\sigma({\bf p},t)C({\bf k}-{\bf p },t);
\end{equation}
the auxiliary-field correlation function $C_\sigma({\bf{k}},t)$ is solution of  the equation
\begin{align}\label{eq:29}
\frac{\partial C_\sigma({\bf{k}},t)}{\partial t}=&  C_\sigma({\bf{k}},t=0)^{-1}
C_\sigma({\bf{k}},t)\nonumber\\
-&\int_0^tdt'D_\sigma({\bf{k}},t-t')\frac{\partial  C_\sigma({\bf{k}},t')}{\partial t'}
\end{align}
with 
\begin{equation}\label{eq:30}
D_\sigma({\bf{k}},t)=-u\int \frac{d^3{\bf q}}{(2\pi)^3}C({\bf q},t)C({\bf k}-{\bf q },t).
\end{equation}
These  equations are  supplemented by  the  initial conditions $C({\bf
k},t=0)=\left( \mu({\bf{k}})-D({\bf{k}},t=0)\right)^{-1}$  and $C_\sigma({\bf
k},t=0)=-\left(  1-D_\sigma   ({\bf{k}},t=0)\right)^{-1}$  which are easily
shown to be identical to the  equilibrium, static SCSA equations first
derived by Bray\cite{AJBa74}.
\section{Transition from ergodic to non-ergodic behavior}
It  is well known that  mode-coupling and related approximations, when
applied  to   glassforming   systems,    may  lead  to     a dynamical
singularity\cite{gotze91,BCKM96}.   This  latter   corresponds  to   a
transition  from an  ergodic  to a  non-ergodic   behavior and is  not
associated  with  any    thermodynamic equilibrium  transition.    For
searching for    such a singularity  in   the above equations,   it is
convenient to Laplace transform Eq.~(\ref{eq:26}), which gives
\begin{equation}\label{eq:31}
\hat{C}({\bf{k}},z)=\frac{-C({\bf{k}},t=0)}
{z-\displaystyle\frac {1} {C({\bf{k}},t=0)(i+\hat{D}({\bf{k}},z))}}.
\end{equation}
An ergodicity-breaking transition is associated with the appearance of
a  non-zero value of the  long-time limit of the correlation function,
$C({\bf{k}},t\to\infty)$;  as    a   result,   in   the    small-$z$   limit,
$\hat{C}({\bf{k}},z) \sim-\frac{C({\bf{k}},t\to\infty)}{z}$     and
$\hat{D}({\bf{k}},z)\sim-\frac{D({\bf{k}},t\to\infty)}{z}$, which when  inserted
in Eq.~(\ref{eq:31}) leads to
\begin{equation}\label{eq:32}
D({\bf{k}},t\to\infty)=\frac{ C({\bf{k}},t\to\infty)}{C({\bf{k}},t=0)(C({\bf{k}},t=0)-C({\bf{k}},t\to\infty))}.
\end{equation}

The kernel  $D({\bf{k}},t\to\infty)$ is  obtained from Eq.~(\ref{eq:27})  for
the      mode-coupling          approximation            and      from
Eqs.~(\ref{eq:28})-(\ref{eq:30}) for  the    dynamical SCSA.

Consider first the equation resulting from the mode-coupling
approximation,
\begin{widetext} 
\begin{equation}\label{eq:33}
\frac{ C({\bf{k}},t\to\infty)}{C({\bf{k}},t=0)-C({\bf{k}},t\to\infty)}=6u^2C({\bf{k}},t=0)\int \frac{d^3{\bf k}_1}{(2\pi)^3}\int \frac{d^3{\bf
k}_2}{(2\pi)^3}C({\bf{k}}_1,t\to\infty )C({\bf{k}}_2,t\to\infty )C({\bf k}-{\bf k
}_1-{\bf k }_2,t\to\infty).
\end{equation}
\end{widetext}

Note that this actually represents a set of coupled equations for the
various ${\bf k}$-modes. The necessary input for solving this equation
is the knowledge of the equilibrium static correlation function
$C({\bf{k}},t=0)$ (see discussion above). We consider here two
standard approximations:

(i) the  Hartree approximation, already presented in
Eqs.~(\ref{eq:5}) and (\ref{eq:6}), and

(ii) the SCSA,  described in the
previous section and leading to
\begin{equation}\label{eq:34}
C^{-1}({\bf{k}},0)=\mu({\bf{k}})+2u\int \frac{d^3{\bf{q}}}{(2\pi)^3}\frac {C({\bf{k}}-{\bf{q}},0)} {1+u\Pi({\bf{q}})}.
\end{equation}
where 
\begin{equation}\label{eq:35}
\mu({\bf{k}})=r_0+k^2+\frac{{Q}}{k^2}+u\int
\frac{d^3{\bf q}}{(2\pi)^3}C({\bf{q}},0)
\end{equation}
 and  
\begin{equation}
\Pi ({\bf{k}})=\int \frac{d^3{\bf q}}{(2\pi)^3}C({\bf{q}},0)C({\bf{k}}-{\bf{q}},0).
\end{equation}
Note that if one neglects the term
$u\Pi({\bf{q}})$   in   Eq.~(\ref{eq:34}),  one   recovers the   Hartree
approximation.

In  both   cases, only   the paramagnetic phase   $<\phi_{\bf k}>=0$,  is
considered.  From  now  on, we  take  $u=1$  (recall that all momentum
integrations are cut-off at $\Lambda=1$).

Other approximations will be discussed in section VI.

\begin{figure}
\centering

\resizebox{8cm}{!}{\includegraphics{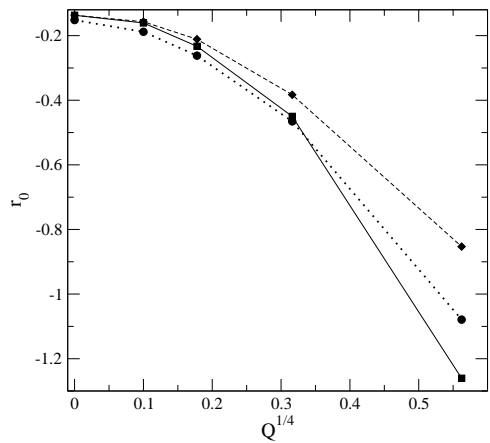}}
\caption{Ergodicity-breaking transition in the $r_0-(Q^{1/4})$ diagram. Dotted
line: mode-coupling approximation with the static Hartree
approximation; dashed line: mode-coupling approximation with the
static SCSA; full line: dynamical SCSA.}
\label{fig:2}

\end{figure}

\begin{figure}
\centering
\resizebox{8cm}{!}{\includegraphics{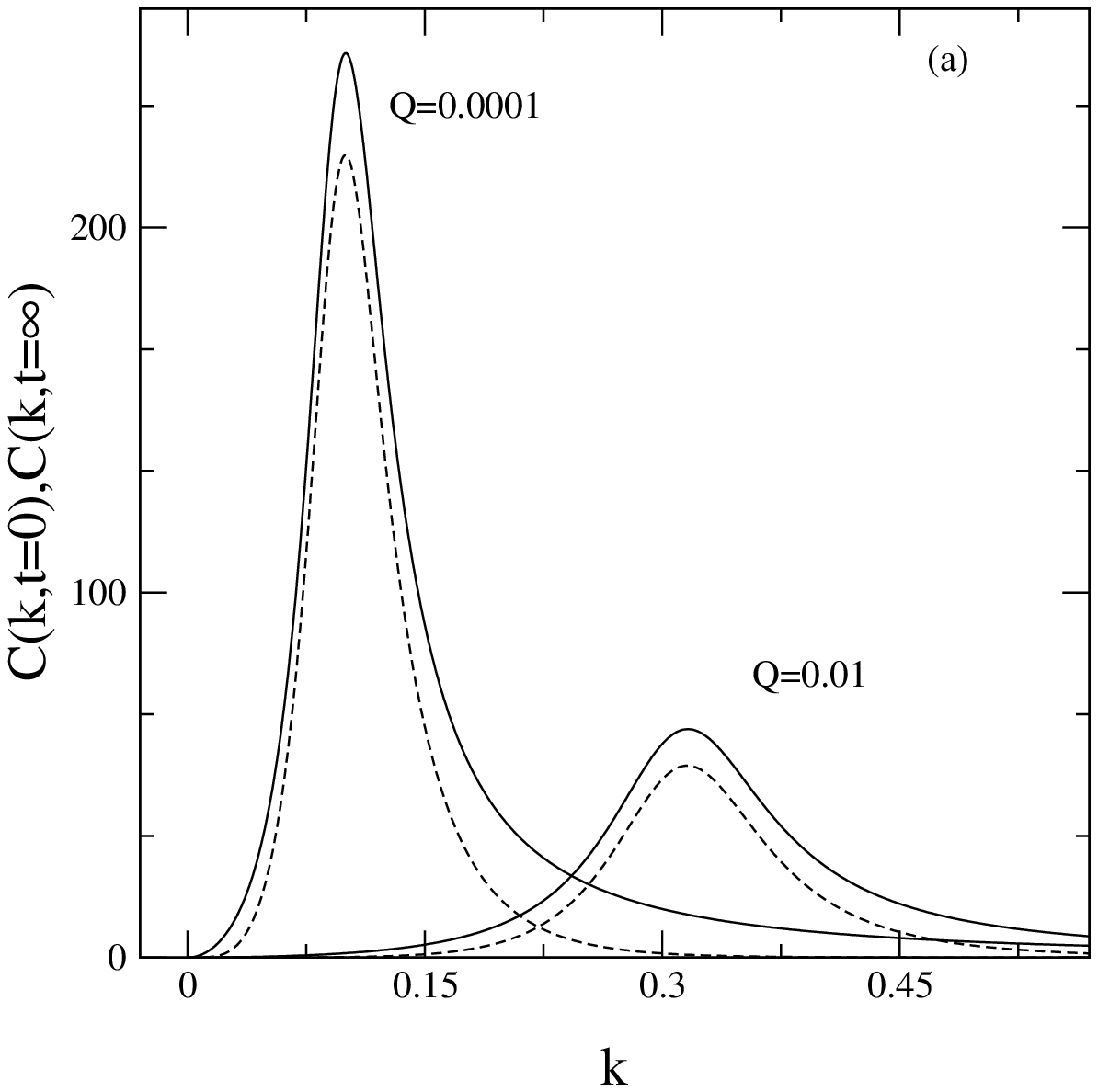}}
\\\resizebox{8cm}{!}{\includegraphics{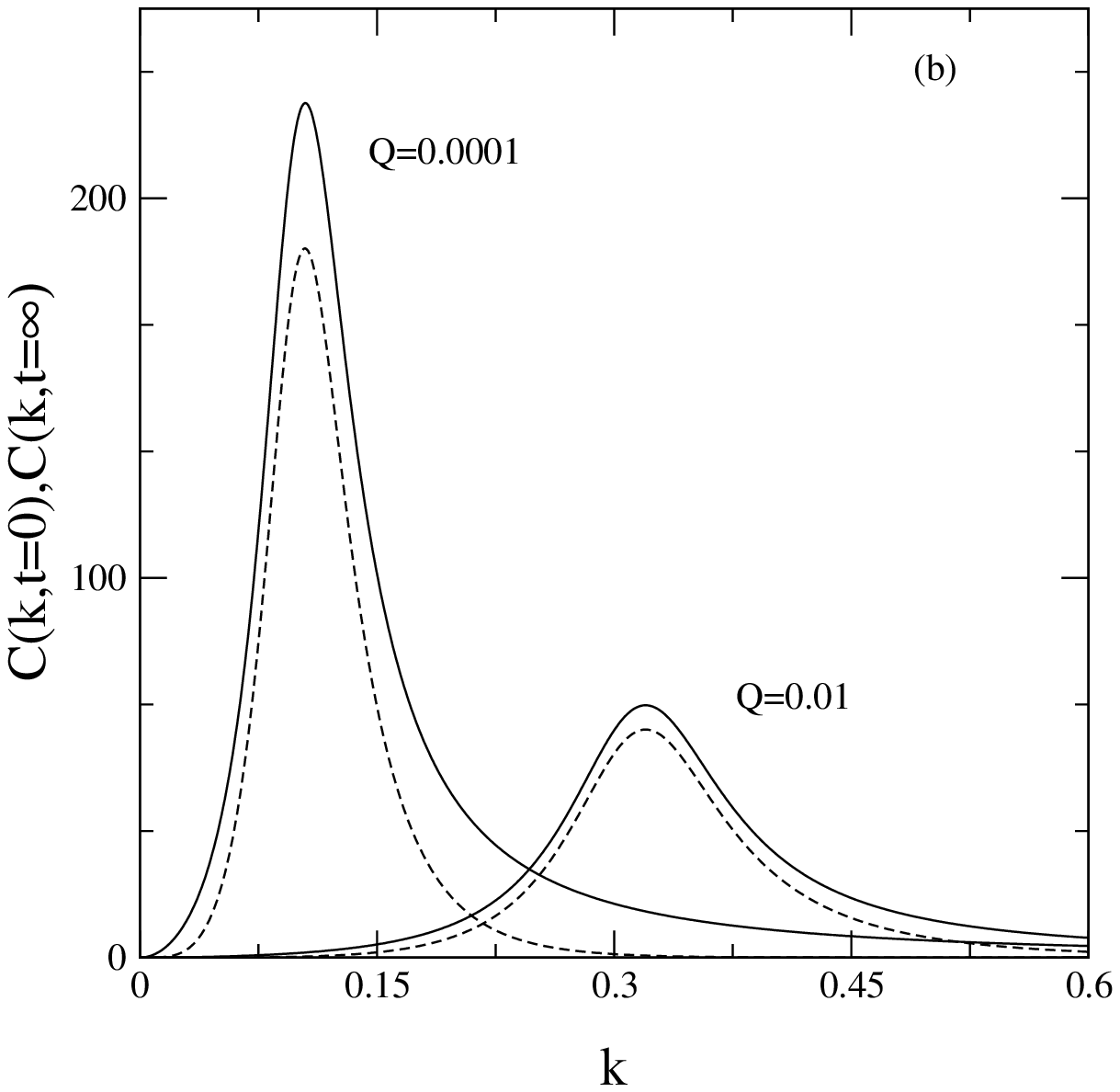}}

\caption{Correlation function $ C({\bf{k}},t)$ at $t=0$  (full line) 
and in the  $t\to\infty$ limit (dashed  line) in the  dynamical mode-coupling
approximation just below the   ergodicity-breaking transition for  two
different  values   of   the  frustration   $Q$: (a)    static Hartree
approximation; (b) static SCSA.}
\label{fig:3}
\end{figure}

\begin{figure}
\centering

\resizebox{8cm}{!}{\includegraphics{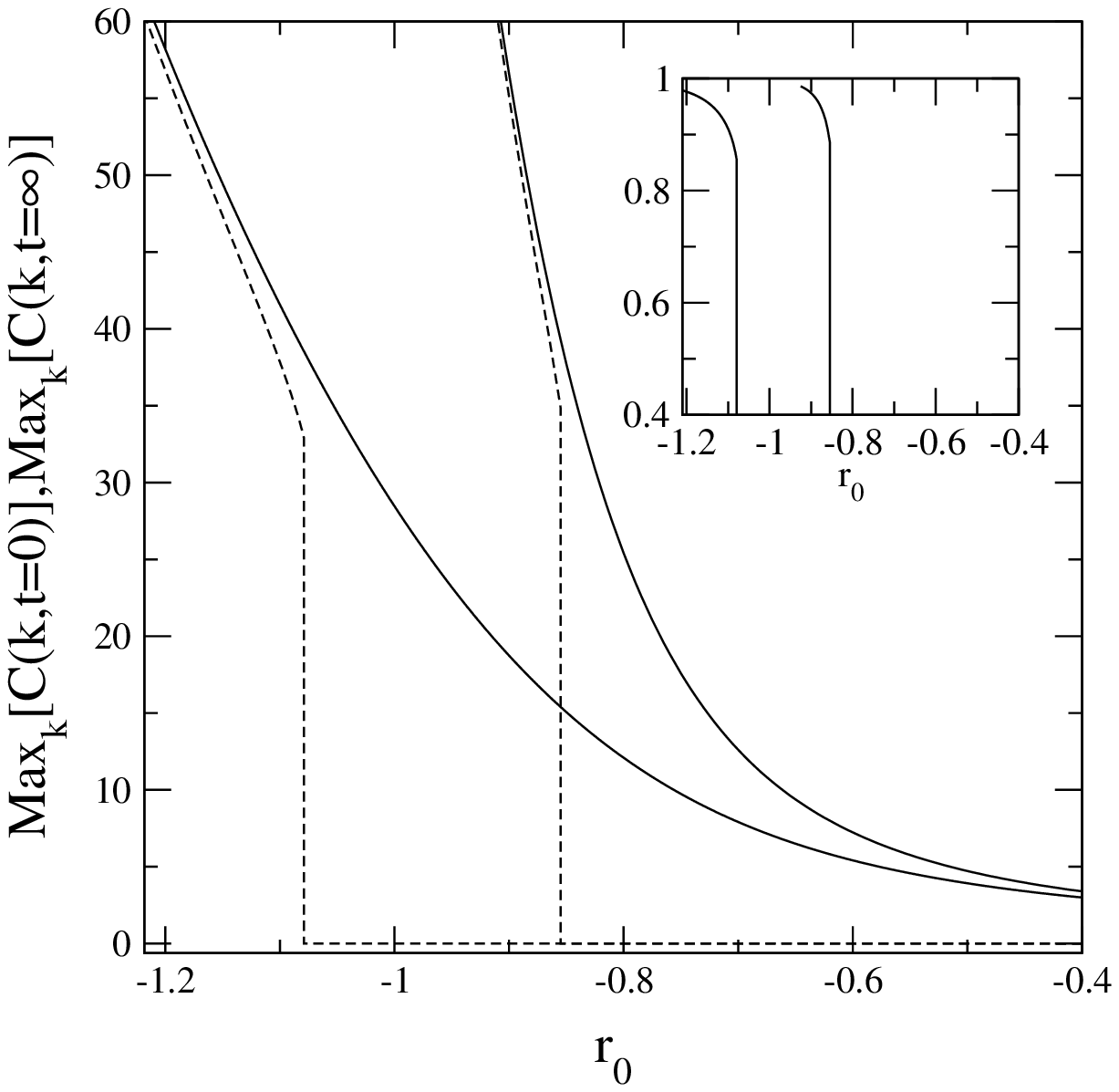}}
\caption{Maximum of the static correlation function ${\rm Max}_{\bf k}\{
C({\bf{k}},t)\}$ versus temperature ($r_0$) for  $t=0$ (full line)  and
$t\to\infty$ (dashed  line) in the dynamic  mode-coupling approximation for a
frustration $Q=0.1$. The left curves correspond  to the static Hartree
approximation and  the right curves  to the static  SCSA.  The jump in
${\rm Max}_{\bf k}\{ C({\bf{k}},t\to\infty )\} $ signals the transition from an
ergodic  (high    temperature) to  a   non-ergodic  (low  temperature)
behavior. The inset shows the normalized non-ergodicity factor ${\rm
Max}_{\bf k}\{ C({\bf{k}},t\to\infty )\}/{\rm Max}_{\bf k}\{ C({\bf{k}},t=0)\} $
versus $r_0$. }\label{fig:4}
\end{figure}

\begin{figure}
\centering

\resizebox{8cm}{!}{\includegraphics{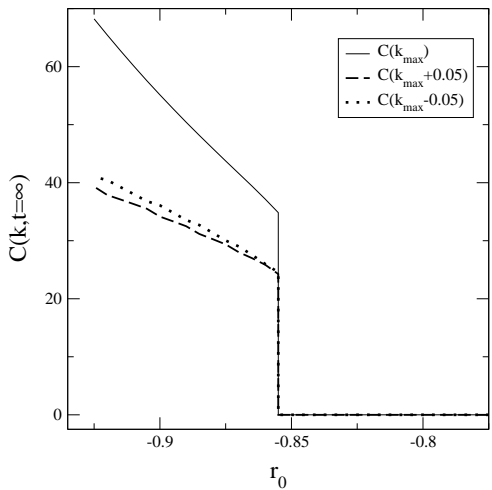}}
\caption{Correlation function $C({\bf{k}},t=\infty)$ versus temperature
($r_0$) for $Q=0.1$ and $3$ different momenta ${\bf{k}}$ corresponding
to the maximum of the function ($k_{\max}$, full line), a higher value
($k_{\max}+  0.05$, dashed line)  and a lower value ($k_{\max}- 0.05$,
dotted   line).  The    results    are  obtained  for   the  dynamical
mode-coupling approximation supplemented  by  the  static  SCSA.  Note
that the jumps occur at the same temperature.}\label{fig:5}
\end{figure}

We have solved the set of  coupled equations, Eq.~(\ref{eq:33}), by an
iterative method.  We find that with the above two approximations, the
mode-coupling approach does lead  to an ergodicity-breaking transition
for the  Coulomb  frustrated model.  When decreasing  the temperature,
i.e.,  the  bare mass $r_0$,  one  reaches  a  point at  which $C({\bf
k},t\to\infty)$  discontinuously jumps to a   non-zero value.  The transition
temperature  increases  as  frustration decreases; it  seems  to reach
continuously, when  $Q\to0$, the  (equilibrium) critical temperature  of
the unfrustrated system, i.e., of the  standard $\phi^4$ theory in either
the Hartree approximation or the SCSA. This behavior is illustrated in
Fig.~\ref{fig:2}.  One   observes  a discrepancy between  the  results
obtained with  the two different approximations   (i) and (ii)  for $
C({\bf{k}},t=0)$, but it  stays within reasonable bounds: the relative
difference   is   about   $20\%$  or  less.     As can    be seen  from
Fig.~\ref{fig:3},     the two  approximations     predict very similar
correlation  functions,   both at   equilibrium  ($t=0$) and   in  the
non-ergodic state  $(t\to\infty)$ when the  temperature is at (or just below)
the dynamical transition.   Roughly speaking, this latter takes  place
when    the maximum of     the  equilibrium  correlation   function  $
C({\bf{k}},t=0)$, a maximum that occurs  for $|{\bf k}|\simeq k_m=Q^{1/4}$,
reaches a given, $Q$-dependent value: this is illustrated for $Q=0.1$
in Fig.~\ref{fig:4}   where  the  dynamical  transition occurs    when
${\rm Max}_{{\bf k}}\{ C({\bf{k}},0)\} \simeq  40$.  As the  maximum increases when
$Q$ decreases slightly more rapidly with the SCSA than with the Hartree
approximation, the    former predicts   a somewhat  higher  transition
temperature than the latter: see Fig.~\ref{fig:4}.  Note that the fact
that the ergodicity-breaking  transition  is driven by the  maximum of
the  ${\bf  k}$-dependent  equilibrium correlation  function  is  well
established in  the context of mode-coupling approaches\cite{gotze91}.
As illustrated in Fig.~\ref{fig:5}, the  transition takes place at the
same temperature for all ${\bf{k}}$-modes.

The transition  form  ergodic to non-erogodic dynamical   behavior can
also   been  studied within   the  dynamical  SCSA.  The corresponding
equations    to    be     solved     are    given     above   and   in
Appendix~\ref{sec:dynam-self-cons}.   A dynamical transition is indeed
found, and the frustration-dependence of the transition temperature is
shown  in Fig.~\ref{fig:2}. The predicted transition  line is not much
different  from    those obtained    with  the   above   mode-coupling
approximations. We show in Appendix~\ref{sec:dynam-self-cons} that the
expressions  for the  correlation function  $C({\bf{k}},t)$ for $t=0$
and $t\to\infty$ derived within the dynamical  SCSA when ergodicity is broken
are identical to those obtained in Refs.\cite{ScWo00,SW201} by using
the  purely thermodynamic analysis based  on the replica formalism and
the  static SCSA.  As  a result, the dynamical  SCSA predicts that the
dynamics  looses  ergodicity  precisely  at   the  point at which   an
exponentially large  number    of metastable   states  occurs  in  the
Schmalian-Wolynes treatment.   Below this temperature,  ergodicity  is
broken: the  fluctuation-dissipation theorem  and the time-translation
invariance no   longer apply,  and    Eqs.~(\ref{eq:26})-(\ref{eq:28})
should be generalized    to    describe the evolution of      two-time
correlation  and   response    functions and   the    associated aging
behavior\cite{BCKM96}.

\begin{figure}
\centering

\resizebox{9cm}{!}{\includegraphics{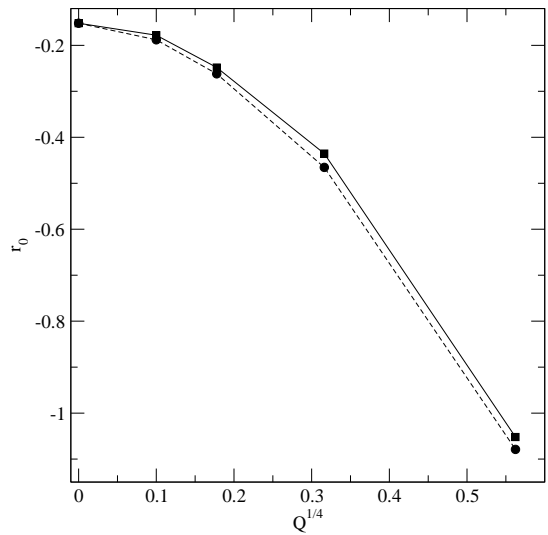}}
\caption{Ergodicity-breaking transition (dynamical mode-coupling approximation
with      static    Hartree    approximation),    full     curve,  and
fluctuation-induced   first-order     transition     (static   Hartree
approximation),       dashed      curve,  in      the    $r_0-Q^{1/4}$
diagram.}\label{fig:6}
\end{figure}

Finally,   it    is instructive   to   compare the   location   of the
mode-coupling-like dynamical transition with  that of the equilibrium,
thermodynamic  transition discussed  in section~\ref{sec:model}.   The
dynamical transition occurs  at a temperature  that is lower than  the
critical temperature of the   unfrustrated system, a  temperature that
was shown in  the Monte Carlo  study of Ref.\cite{GrTV01}  to mark the
onset  of  fragile glassforming behavior;   it seems   to occur  at  a
temperature   close  to  that of the   fluctuation-induced first-order
transition from   the paramagnetic to the  modulated  phases.  This is
illustrated in   Fig.~\ref{fig:6}  where we  display  the  first-order
transition    obtained  within    the    Hartree  approximation   (see
section~\ref{sec:model}  and  Fig.~\ref{fig:1})   and    the dynamical
transition   obtained   within  the   mode-coupling      approximation
supplemented by the static Hartree  approximation (as discussed above,
the other predictions are quite close to this latter).

Actually at small  $Q$'s the dynamical  transition  appears even below
the temperature of  the equilibrium first-order  transition: in such a
case, the dynamical transition takes   place in the {\it  supercooled}
paramagnetic (``liquid'') regime, a regime that appears because of the
first-order nature of the transition  to the modulated phases and that
can be  described  by the    Hartree approximation.  As
discussed  in  section~\ref{sec:model},   the   paramagnetic phase  is
(meta)-stable  at  all finite  temperatures within this approximation,
and the  equilibrium correlation length is  therefore  finite in all the
region where the dynamics are studied.
\section{Evolution with time of the correlation function}
\begin{figure}
\centering

\resizebox{8.5cm}{!}{\includegraphics[angle=270]{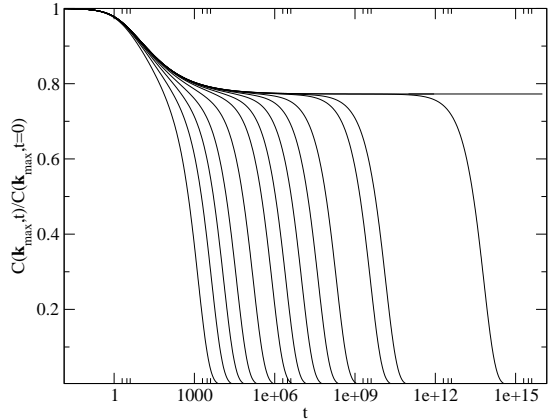}}
\caption{Time dependence of the correlation function
$C({{\bf{k}}_{\max}},t)$ for  $Q=0.1$ as predicted by the mode-coupling
approximation       (supplemented     by     the     static    Hartree
approximation).   Curves from left to  right  correspond to decreasing
temperatures  : $r_0=-1.05$,   $-1.06$, $-1.065$, $-1.068$,  $-1.070$,
$-1.071$,          $-1.07135$,$-1.07155$,$-1.07165$,        $-1.0717$,
$-1.07173$,  $-1.071734$,  $-1.0717379$, and  $-1.071738$ .The  dynamical transition is  at
$r_0\simeq -1.0717379\ldots$}\label{fig:7}
\end{figure}

We have also solved  the full set  of coupled equations describing the
time   evolution of      the  equilibrium    correlation function    $
C({\bf{k}},t)$      in          the    mode-coupling    approximation,
Eqs.~(\ref{eq:26})-~(\ref{eq:27}).  (The    algorithm is described  in
Refs\cite{FGHL91,K00}.)  For the input quantity, $C({\bf{k}},t=0)$, we
have used  the  Hartree  approximation.  The  results   are  shown  in
Fig.~\ref{fig:7}   for  the time-dependent  correlation  function at a
momentum $k_{\max}\simeq k_m=Q^{1/4}$ that corresponds to the maximum value
of  the  function; curves for the   frustration parameter  $Q=0.1$ and
several temperatures (i.e., several values of the bare mass $r_0$) are
shown.  One observes a behavior typical of the mode-coupling equations
with a so-called  $B$-type  transition\cite{gotze91} as those  used to
describe glassforming liquids\cite{gotze91}  and those  describing the
dynamics  of     a    class     of  mean-field      generalized   spin
glasses\cite{BCKM96}.

At high temperature, the correlation function decays  in one step, but
as  temperature is   lowered a second   relaxation step  appears, that
becomes slower and slower  so that a  plateau develops between the two
relaxation steps. When temperature is further decreased, one reaches a
point at  which the slow  (``$\alpha$'')  relaxation time  diverges. The
correlation  function no   longer decays  to  zero, but  stays at  the
plateau   value.   Below this   point,  ergodicity is   broken and the
mode-coupling equations derived  under  the condition  of  equilibrium
(with the   fluctuation-dissipation  theorem and  the time-translation
invariance) are no longer valid.

\begin{figure}
\centering
\resizebox{8cm}{!}{\includegraphics{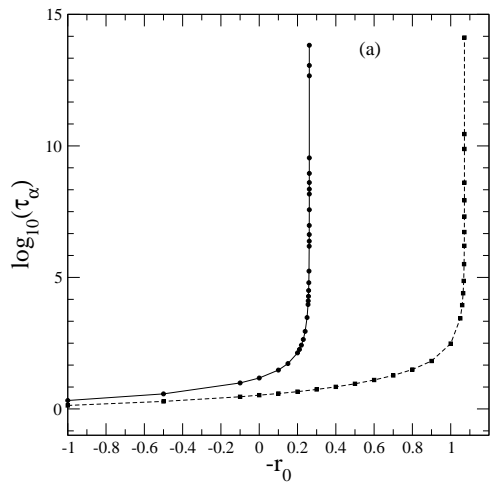}}\\

\resizebox{8cm}{!}{\includegraphics{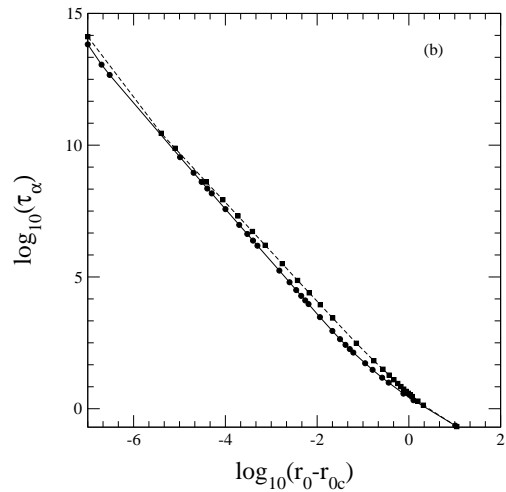}}
\caption{Temperature dependence of the $\alpha$-relaxation time $\tau_\alpha $
obtained from the dynamical mode-coupling  prediction (with the static
Hartree approximation) for the time  dependent correlation function  $
C({\bf{k}_{\max}},t)$.   $\tau_\alpha$  is   defined  as the  time   at  which
$C({\bf{k}_{\max}},t)=0.1$.  Two frustrations, $Q=0.1$ and $Q= 0.001$,
are  shown. (a) $\log_{10}(\tau_\alpha)$ versus  $-r_0$;  the  left and  right
curves correspond     to $Q=0.1$ and  $Q=  0.001$   respectively.  (b)
$\log_{10}(\tau_\alpha)$  versus $\log_{10}(r_0-r_{0c}(Q))$, where $r_{0c}(Q)$
is the ergodicity-breaking transition point. }
\label{fig:8}
\end{figure}

In the vicinity of the dynamical transition, various scaling laws are
observed, and the slow (``$\alpha$'') relaxation time diverges as a power law,
\begin{equation}
\tau_\alpha(Q,T)\sim (T-T_c)^{-\gamma},
\end{equation}
where   the   exponent $\gamma\simeq  1.85-1.89$  is    weakly  dependent on the
frustration parameter, provided  $Q>0$.  (For $Q=0$, the  system shows
standard critical slowing down with  $\tau(T)\sim (T-T_c)^{-z\nu }$, where $z$
is the dynamical  exponent  and $\nu$  the (static) correlation   length
exponent\cite{HH77}.) For illustration,  we have plotted the logarithm
of the $\alpha-$ relaxation time versus temperature $r_0$ for two different
frustrations in Fig.~\ref{fig:8}.

\section{Sensitivity of the results to the approximation scheme}
We  have  already  mentioned  (see Ref.~\cite{endnote3}) that   enough  of the
non-linearities of the original dynamical equation or of the equations
in replica space must be kept in any approximate treatment in order to
find non-trivial phenomena such as  ergodicity breaking and appearance
of  an  exponentially  large number of   metastable  states.  For this
reason, the  dynamical as  well as the   replica-space 
Hartree  approximations  are unable to  generate  such phenomena.  One
must therefore consider    improved resummation schemes  such as   the
mode-coupling approximation  and the SCSA\cite{endnote4}.

\begin{figure}
\centering

\resizebox{8cm}{!}{\includegraphics{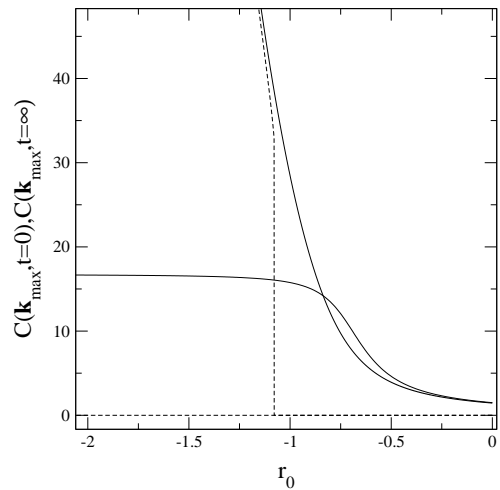}}
\caption{Maximum of the 
correlation function $C({\bf{k}_{\max}},t)$ versus temperature ($r_0$)
for $t=0$  (full  lines) and  $t\to\infty$  (dashed  line)  in the  dynamical
mode-coupling  approximation.   The    value  of   the  frustration is
$Q=0.1$.  The  two  upper curves   correspond to  the  static  Hartree
approximation that  predicts  an ergodicity  breaking  transition. The
lower curve that saturates at low temperatures corresponds to the less
renormalized version  of  the SCSA (Eq.~(\ref{eq:36})):  no  dynamical
transition is observed in this case.}
\label{fig:9}

\end{figure}

The additional point we would like to make here is that even in the
mode-coupling approximation, the results are somewhat dependent upon
the supplementary approximation which is made to describe the static
properties of the system  entering as an input in the dynamical
equation. It is well known, and was recalled above, that the location,
or even the existence of an ergodicity-breaking transition is sensitive
to the amplitude of the peak in the equilibrium (static) correlation
function $C({{\bf{k}},t=0})$. We have seen that the Hartree and SCSA give
slightly different, but compatible results. If one uses instead a
somewhat less renormalized version of the SCSA with $\mu ({\bf{k}})$ in
Eq.~(\ref{eq:35}) now defined with the Hartree and not the full
correlation function, i.e.,
\begin{equation}\label{eq:36}
\mu({\bf{k} })=r_0+k^2+\frac{{{Q}}}{k^2}+u\int\frac{ d^3{\bf q}}{(2\pi)^3}\frac{1}{\mu({\bf{q}})},
\end{equation}
a different behavior is obtained. As shown in Fig.~\ref{fig:9}, the
maximum of the static correlation function appears to saturate, as one
lowers the temperature, to a value that is too small to trigger a
breaking of ergodicity. 

Finally,  considering  the   static   analog  of the   mode-coupling
approximation to  compute  $C({{\bf{k}},t=0})$  (see Eq.~(\ref{eq:25})
and below), i.e.,
\begin{align}\label{eq:37}
 C^{-1}({\bf{k}},t=0)=& r_0+k^2+\frac{Q}{k^2}+3u\int \frac{d^3{\bf q}}{(2\pi)^3} 
C({\bf{q}},t=0)\nonumber\\
-&D({\bf{k}},t=0).
\end{align}
where $D({{\bf{k}},t=0})$ is given by Eq.~(\ref{eq:27}) 
leads
to  a situation  in which  the limit of   stability  (spinodal) of the
paramagnetic phase is   reached at a  finite  temperature,  before the
occurrence of an ergodicity-breaking transition.

The   validity of these  various   approximations should  of course be
checked by performing a computer simulation of the model. However, one
can tentatively conclude  from  the above  exercise that  despite  the
formal similarity between the dynamics and the statics that comes from
using  the  Martin-Siggia-Rose    functional    formalism\cite{MSR78},
different levels   of approximation may  be  required to  describe the
dynamical and the static properties of the Coulomb frustrated model.
\section{Conclusion}
We  have  studied the Langevin  dynamics of   the soft-spin, continuum
version  of the Coulomb   frustrated Ising ferromagnet.  By using  the
dynamical    mode-coupling   approximation,   coupled  with reasonable
approximations for  describing   the equilibrium   static  correlation
function,  and the  dynamical self-consistent screening approximation,
we have   found that the  system's dynamics  display a transition from
ergodic  to  non-ergodic behavior, similar  to   that obtained  in the
idealized mode-coupling theory of glassforming   liquids\cite{gotze91}
and in the mean-field  generalized spin-glasses with  one-step replica
symmetry breaking\cite{KT87,BCKM96}.  This  transition  occurs in  the
paramagnetic phase, either in the stable or the metastable region.  It
is  related  to the  emergence of   an exponentially large  number  of
metastable  states  found by  a   purely static replica  approach: the
system looses ergodicity because it gets  trapped in free-energy minima
separated   from  each   other  by  infinite   barriers.   This  whole
description,   as can   be inferred  from    the very  nature  of  the
approximations that  amount  to  partial resummations of  perturbative
expansions, has  a mean-field character: thermally activated processes
are completely  ignored.  The predicted  singularity is ``avoided'' in
the true dynamics  of the system and  it remains to  be seen, e.g., in
computer simulations of Coulomb frustrated models, what signatures may
still be observed  in the time  evolution of the correlation function.
As for   describing the activated  processes, other,  non-perturbative
approaches,   such as  the    phenomenological     frustration-limited
domain\cite{KKZNT95}  and entropic-droplet pictures\cite{KTW89},
must be used.

\appendix
\section{Fluctuation-induced first order transition}\label{sec:static-first-order}
In this appendix, we   calculate  the temperature of  the  equilibrium
transition between the paramagnetic  phase  and phases with  spatially
modulated order within the  self-consistent Hartree approximation. The
derivation  given    below   closely  follows    Brazovskii's original
treatment\cite{B75}.

The starting point is the Hamiltonian in Eq.~(\ref{eq:2}) augmented by
the introduction  of spatially varying  external fields $h_{{\bf{k}}}$
that are  linearly  coupled to the scalar   field $\phi_{{\bf{k}}}$. As a
result,  $\phi_{{\bf{k}}}$ is  now the    sum  of an average   component,
$m_{{\bf{k}}}=  <\phi_{{\bf{k}}}>$,  and a fluctuation     $\psi_{{\bf{k}}}=
\phi_{{\bf{k}}}-m_{{\bf{k}}}$. The self-consistent  Hartree approximation
is then equivalent to   a gaussian variational approximation  for  the
fluctuations. The resulting equation of state reads
\begin{widetext}
\begin{align}\label{eq:38}
h_{{\bf{k}}}=&\left(r_0+k^2+\frac{Q}{k^2}\right)m_{{\bf{k}}}
+u\int \frac{d^3{\bf k}_1}{(2\pi)^3}\int\frac{d^3{\bf k}_2}{(2\pi)^3}\left(m_{{\bf{k}_1}}
m_{{\bf{k}_2}}+3C({\bf{k}_1},{\bf{k}_2})\right)m_{{\bf{k-k}_1-\bf{k}_2}},
\end{align}
where the connected correlation function
$C({\bf{k}},{\bf{k}'})=<\psi_{{\bf{k}}}\psi_{{\bf{k}}'}>$, is obtained
self-consistently by solving
\begin{equation}\label{eq:39}
C^{-1}({\bf{k}},{\bf{k}'})=\left(r_0+k^2+\frac{Q}{k^2}\right)\delta({\bf{k}}+{\bf{k}'})+
3u\int \frac{d^3{\bf q}}{(2\pi)^3}\left(m_{{\bf{q}}}m_{{\bf{k+k'-q}}}+C({\bf{q}},{\bf{k}}+{\bf{k}'}-{\bf{q}})\right),
\end{equation}
\end{widetext}
together with the inversion formula
\begin{equation}\label{eq:40}
\int \frac{d^3{\bf q}}{(2\pi)^3}C^{-1}({\bf{k}},{\bf{q}})C({\bf{q}},{\bf{k}'})=\delta ({\bf{k}'}-{\bf{k}}).
\end{equation}
In the  paramagnetic    phase     when  all    $h_{{\bf{k}}}$'s,    and
therefore $m_{{\bf{k}}}$'s,    are  equal  to      zero,   Eqs.~(\ref{eq:39}) and
(\ref{eq:40})        reduce         to      Eq.~(\ref{eq:5})      with
$C({\bf{k}},{\bf{k}'})=C({\bf{k}})\delta ({\bf{k}}+{\bf{k}'})$.

In the vicinity of the transition between paramagnetic and modulated
phases and for small enough frustration $Q$ ($Q<<1$), the modulated order is
one-dimensional and characterized by a wave-vector ${\bf{k}}_m$ with
$|{\bf{k}}_m|=k_m=Q^{1/4}$. It is then sufficient to consider
\begin{equation}\label{eq:41}
h_{{\bf{k}}}=\tilde{h}\left(\delta({\bf{k-k}_m})+\delta({\bf{k+k}_m})\right)
\end{equation}
and
\begin{equation}\label{eq:42}
m_{{\bf{k}}}=\tilde{m}\left(\delta({\bf{k-k}_m})+\delta({\bf{k+k}_m})\right).
\end{equation}
In this region, the fluctuations of wave-vector ${\bf{k}}$ with
$|{\bf{k}}|=k_m$ are dominant, and Brazovskii\cite{B75} has shown that the
effect of the off-diagonal terms with ${\bf{k}}\neq {\bf{k}}'$ could be
neglected in the correlation function. As a result, 
\begin{equation}\label{eq:43}
C({\bf{k}},{\bf{k}'})\simeq \frac{\delta({\bf{k+k}'})}{r+k^2+\frac{Q}{k^2}},
\end{equation}
where the renormalized mass in a  phase characterized by
Eqs.~(\ref{eq:41},\ref{eq:42}), is given by:
\begin{align}\label{eq:44}
r=&r_0+3u\int\frac{d^3{\bf
k}}{(2\pi)^3}\left(\frac{1}{r+k^2+\frac{Q}{k^2}}+|m_{{\bf{k}}}|^2\right)
\nonumber\\=&r_0+3u\int\frac{d^3{\bf
k}}{(2\pi)^3}\frac{1}{r+k^2+\frac{Q}{k^2}}+6u|\tilde{m}|^2.
\end{align}
By introducing Eqs.~(\ref{eq:41})-~(\ref{eq:44}) in Eq.~(\ref{eq:38})
 and recalling that $k_m=Q^{1/4}$, one obtains the following equation
of state: 
\begin{align}
\tilde{h}=&\left(r_0+2Q^{1/2}\right.\nonumber\\\label{eq:45}
+&\left.3u\int\frac{d^3{\bf
k}}{(2\pi)^3}\frac{1}{r+k^2+\frac{Q}{k^2}}+3u|\tilde{m}|^2\right)\tilde{m}\nonumber\\
=&\left(r+2Q^{1/2}-3u|\tilde{m}|^2\right)\tilde{m}.
\end{align}
In the  Hartree approximation, and below some  temperature, there is a
coexistence of the paramagnetic phase and the modulated phase. In zero
field ($\tilde{h}=0$), the  former  is characterized  by $\tilde{m}=0$
and the  latter  by $\tilde{m}\neq 0$,  where $\tilde{m}$  is solution of
$\left(r+2Q^{1/2}-3u|\tilde{m}|^2\right)=0$.   The  transition  point,
which is then associated with a first-order transition, is obtained as
the temperature  at  which the free-energies of   the  two phases  are
equal. Following  Brazovskii\cite{B75}, it is  convenient to calculate
directly the free-energy  difference $\Delta F(r_0)$  between the modulated
($\tilde{m}\neq 0$) and the paramagnetic   ($\tilde{m}= 0$) phases at   a
given temperature $r_0$ from the following expression:
\begin{align}\label{eq:46}
\Delta F=&\int_{0}^{\tilde{m}}  d\tilde{m}'\frac{\partial F}{\partial \tilde{m}'}=2\int_{0}^{\tilde{m}}d\tilde{m}'\tilde{h}(\tilde{m}')
\end{align}
where $\tilde{h}_{\tilde{m}'}$  is      given  by  Eq.~(\ref{eq:45}). One can
change the integration variable from $\tilde{m}'$ to $r'$ with
$r'(\tilde{m}')$ solution of Eq.~(\ref{eq:44}). After some algebra,
Eq.~(\ref{eq:46}) can be recast as
\begin{widetext}
\begin{align}\label{eq:47}
u\Delta F=&\int_{r(\tilde{m}=0)}^{r(\tilde{m})}  d\tilde{r}'\left(\frac{r'+r_0}{2}+2Q^{1/2}   +\frac{3u}{4\pi^2} \int
dk\frac{k^2}{r+k^2+\frac{Q}{k^2}} \right) \left(\frac{1}{6}+\frac{u}{4\pi
^2}\int dk\frac{k^2}{\left(r'+k^2+\frac{Q}{k^2}\right)^2} \right),
\end{align}
\end{widetext}
where   $r(\tilde{m}=0)$   is  solution    of  Eq.~(\ref{eq:44})   with
$\tilde{m}=0$,   i.e.,   of Eq.~(\ref{eq:6}),  and  $r(\tilde{m})$ and
$\tilde{m}$    are   solutions  of     the  two   coupled   equations,
Eq.~(\ref{eq:44}) and $\left(r+2Q^{1/2}-3u|\tilde{m}|^2\right)=0$.  By
solving  Eq.~(\ref{eq:47})  numerically for several  values  of $Q<<1$
(and for $u=1$),  we have found  that the sign of $\Delta  F$ changes  at a
finite value of $r_0$ that  marks the   first-order
transition between  paramagnetic  and modulated phases. The  result is
shown in Fig.~\ref{fig:1}. The transition being second-order in the
mean-field approximation, it is then driven first-order by the fluctuations.

\section{Ergodicity breaking in the Dynamical SCSA}\label{sec:dynam-self-cons}
One can see from Eqs.~(\ref{eq:28})-~(\ref{eq:30}) that ergodicity
breaking requires that both $ C({\bf{k}},t)$ and  $ C_\sigma ({\bf{k}},t)$, and
as a consequence  $ D({\bf{k}},t)$ and $ D_\sigma ({\bf{k}},t)$, go to non-zero
values in the limit $t\to\infty$. From Eq.~(\ref{eq:32})), one obtains
\begin{equation}\label{eq:48}
  C({\bf{k}},t\to\infty )=\frac{ D({\bf{k}},t\to\infty ) C({\bf{k}},0)^2}
{1+D({\bf{k}},t\to\infty )C({\bf{k}},0)}.
\end{equation}
A similar expression can be derived for $ C_\sigma ({\bf{k}},t\to\infty )$ by
first Laplace transforming Eq.~(\ref{eq:29}),
\begin{equation}\label{eq:49}
  \hat{C}_\sigma ({\bf{k}},z )=\frac{ -C_\sigma ({\bf{k}},0 )}
{z-\displaystyle\frac{1}{C_\sigma ({\bf{k}},0)(i +\hat{D}_\sigma({\bf{k}},z))}}\,,
\end{equation}
and by looking for the dominant behavior in the small-$z$ limit,
$\hat{C}_\sigma ({\bf{k}},z )\sim -C_\sigma ({\bf{k}},t\to\infty  )/z$, 
$\hat{D}_\sigma ({\bf{k}},z )\sim -D_\sigma ({\bf{k}},t\to\infty  )/z$; one finally gets
\begin{equation}\label{eq:50}
C_\sigma ({\bf{k}},t\to\infty  )=\frac{-D_\sigma ({\bf{k}},t\to\infty  )C_\sigma ({\bf{k}},0 )^2}
{1-D_\sigma ({\bf{k}},t\to\infty  )C_\sigma ({\bf{k}},0 )}.
\end{equation} 
By introducing the time-dependent polarization 
\begin{equation}
\Pi({\bf{k}},t )=\int\frac{d^3{\bf{q}}}{(2\pi)^3}C({\bf{q}},t )C({\bf{k-q}},t
),
\end{equation}
one can express the memory kernel $D_\sigma ({\bf{k}},t  )$ given in
Eq.~(\ref{eq:30}) as
\begin{equation}\label{eq:51}
D_\sigma ({\bf{k}},t  )=-u\Pi({\bf{k}},t ),
\end{equation}
so that the $t=0$ and $t\to\infty$ values of $C_\sigma ({\bf{k}},t  )$ (given
below Eq.~(\ref{eq:30}) and in Eq.~(\ref{eq:50}), respectively) can be
written as
\begin{align}\label{eq:52}
C_\sigma ({\bf{k}},0  )=&\frac{-1}{1+u\Pi ({\bf{k}},0  )}\\
C_\sigma ({\bf{k}},t\to\infty  )=&\frac{-u\Pi ({\bf{k}},t\to\infty )C_\sigma ({\bf{k}},0
)^2}
{1+u\Pi ({\bf{k}},t\to\infty )C_\sigma ({\bf{k}},0  )}.
\end{align}

Recalling that  $C ({\bf{k}},0  )=(\mu ({\bf{k}} )-D({\bf{k}},0 ))^{-1}$
with                                                                $\mu
({\bf{k}})=r_0+k^2+Q/k^2+u\int\frac{d^3{\bf{q}}}{(2\pi)^3}C({\bf{q}},0   )$
and that $D ({\bf{k}},t )$ is  given by Eq.~(\ref{eq:27}), one obtains
with Eqs.~(\ref{eq:48}), (\ref{eq:51}) and  (\ref{eq:52}) a closed set
of  equations   that   determines  the  non-ergodicity   parameter  $C
({\bf{k}},t\to\infty )$. If one changes  the notations from $C ({\bf{k}},0 )$
and    $C     ({\bf{k}},t\to\infty      )$  to    $\cal{G}({\bf{k}})$     and
$\cal{F}({\bf{k}})$, from $-D({\bf{k}},0 )$ and $-D({\bf{k}},t\to\infty )$ to
$\Sigma ({\bf{k}})$ and $\Sigma_\sigma   ({\bf{k}})$,  from $-uC ({\bf{k}},0  )$  and
$-uC_\sigma ({\bf{k}},t\to\infty )$   to ${\cal D}_{\cal G}({\bf{k}})$ and  ${\cal
D}_{\cal F}({\bf{k}})$, one can easily check  that the above equations
are identical to those   obtained in Refs\cite{ScWo00,SW201} with  the
replica formalism and the static SCSA.

%\bibliography{fld}

\begin{thebibliography}{10}
\bibitem{endnote1} One should note
that $T_K$ is very close of $T_A$,  the relative difference being less
than $10\%$ in the small frustration range, $Q\leq 0.1$.
\bibitem{endnote2}
For  a field theory with $n$-component
fields with $O(n)$   symmetry, the  Hartree approximation takes   into
account all diagrams that survive in the $n\to\infty $ limit whereas the SCSA
includes in a self-consistent manner the leading $1/n$ corrections.  In
the present case, one is  of course interested  by $n=1$.
\bibitem{endnote3}
Contrary to those two  approximations
that retain some of the non-linearities of the dynamics, the dynamical
counterpart of the Hartree  approximation  leads to an  essentially  linear
equation, and it is not further considered in this work.
\bibitem{endnote4} 
Note that  the same
considerations apply to  the replica approach  of the glassy state of
simple liquids,  where  non-trivial behavior   can be  found  with the
hypernetted  chain approximation, but    not with the  ``too  linear''
Percus-Yevick and mean-spherical approximations\cite{MP99}.
\end{thebibliography}

\end{document}